\documentclass[12pt,A4]{article}
\usepackage{epsfig}
\usepackage{array}
\usepackage{amsmath}
\usepackage{amssymb}

\newcommand{\m}{\mathbf}

\newcommand{\be}{\begin{equation}}
\newcommand{\ee}{\end{equation}}
\newcommand{\bea}{\begin{eqnarray}}
\newcommand{\eea}{\end{eqnarray}}
\newcommand{\oh}{\frac{1}{2}}

\begin{document}

\title{THE THEORY OF THE BOHR---WEISSKOPF EFFECT IN THE HYPERFINE STRUCTURE}
\author{ F. F. Karpeshin \\ Mendeleev All-Russian Research
Institute of Metrology, \\ Saint-Petersburg, Russia \\
\bigskip \\
and \\ M.B.Trzhaskovskaya \\
Petersburg Nuclear Physics Institute, Kurchatov Research Center, \\ Gatchina,
Russia }
\maketitle

\abstract{For twenty years research into the anomalies in the HF spectra was going in a wrong direction in fighting the  Bohr---Weisskopf  effect. As  way out, we propose the model-independent way, which enables  the
nuclear radii and their moments to be obtained from the hyper-fine splitting. The way is based on
 analogy of HFS to internal conversion coefficients, and
the Bohr---Weisskopf  effect --- to the anomalies in the internal conversion coefficients. 
It is shown that the parameters which can be extracted from the data are the even nuclear moments of the magnetization distribution. 
The radii $R_2$ and (for the first time) $R_4$ are obtained in this way by analysis of the experimental HFS for the H- and Li-like ions of $^{209}$Bi. 
The critical prediction is made concerning the HFS for the $2p_\oh$ state.
The moments may be determined in this way only if the higher QED effects are properly taken into account. Therefore, this set of the parameters  
 form a basis of a strict QED test. Experimental recommendations are given, aimed at retrieving data on the HFS values for a set of a few-electron configurations of various  atoms. }

\newpage

\section{Introduction}

Hyperfine structure e.g. of the $D_1$ line of the $^{133}$Cs atom or that of the $1s-2s$ transition of the hydrogen atom play an important role in construction of the atomic clock (e.g. \cite{hf}). Experiments with the hyperfine structure allowed one to obtain more accurate value of the hyperfine structure constant $\alpha$ --- one of the fundamental constants. Worthy of notion is also the project of composing  a reference point of frequency, founded on the few-electronvolt isomeric state of the $^{229}$Th nuclide \cite{reller}, with unprecedented systematic shift suppression, allowing for the atomic clock performance with a total fractional
inaccuracy approaching  10$^{-19}$ -- 10$^{-21}$. Hyperfine structure is of especially great importance for the proper design of such clock \cite{peik}.  At the same time, proper account of the hyperfine structure may radically change the estimated lifetime of the nuclear isomeric states \cite{wars}.

       From the viewpoint of theory, the  hyperfine structure arises due to interaction of the electron magnetic moments with the nuclear magnetic moments. The interaction leads to a hyperfine splitting (HFS). That also depends on the
electron density near the nucleus and other properties of the
electron shell. In the first approximation, both nuclear and atomic  factors act
independently, which manifests itself in the \emph{factorization}
of the  nuclear and electron parameters in the related formulae.
However, precision measurements showed that the factorization is
violated for the reason of the finite nuclear size. First of all,
this is the Bohr-Weisskopf effect: finite distribution of the magnetic
currents inside the nucleus. Study of the magnetic anomalies
offers a way of  pursuing the change of the effective nuclear radii within
isotopic chains with addition of neutrons to the nucleus, which
information is of primordial interest.
That explains why a specific problem of  a magnetic anomaly of the
hyperfine splitting of the atomic structure became very topical for the past time \cite{AD,barza,shab1,shab2}. Nuclear-optical methods became very
important in the research into the properties of rare or
radioactive nuclides.

       From the theoretical point of view, description of the hyperfine
splitting is a challenging problem, as it needs a calculation of
the atomic wavefunctions with the accuracy of $\sim 10^{-4}$
\cite{AD,kara}. A convenient way of comparison of the data to theoretical
calculations is offered by investigation of the HFS in
few-electron atoms: H-, Li- or B-like ions.  In these cases, the
electronic wavefunctions can be calculated more reliably.
Moreover, comparison to the theory can be used as a test for our
capability of description of both the electronic structure of the
atoms, as well as QED. In this aspect, the problem of the
Bohr-Weisskopf effect in appearance of the magnetic  anomaly
 is put forward in the first place. Existing method of its
description through construction of  the specific differences is however far from being appropriate and sufficient
for this role. It is erroneous in principle, as it is clear from the internal conversion theory and is shown in the present paper.

    An adequate way of solving this problem was proposed in refs.
\cite{HF,echa,book}.  An analogy was noted between the internal
conversion coefficients (ICC) and the HFS. It was shown that the
HFS can be calculated as the limiting case of the ICC at the
transition energy $\omega \to 0$.  Test calculations were made for
the H-like ions of $^{209}$Bi.  Our present purpose is to show
that the application of the classical theory of  anomalous internal conversion 
\cite{church,anom} gives us the proper tool for adequate description of the Bohr-Weisskopf effect, also offering the way of extracting the unique information about the nuclear
radii. To this end,  we consider the HFS for
the $s$-electronic states. In this case, first, the effect is
maximal, and second, the quadrupole effect of the nuclear deformation  is absent.

\section{Physical premises}

In the case of the $M1$ transitions, in the long-wavelength limit, the
electromagnetic interaction of the nuclei are determined by the
single formfactor, or the reduced current of the nuclear transition
$J(R)$. We define it according to \cite{rose}:
\be \int\left( \m
J_0(\m R)\m T^{(0)^*}_{LM}(\hat R)\right)\,d\Omega_R=
\frac{C(I_fM_fLM|I_iM_i)} {\sqrt{2I_i+1}}iJ(R)\,.
 \label{2} \ee 
Here $\m T_{LM}^{(\lambda)}$ are vector
spherical harmonics:  
\be \m T_{LM}^{(\lambda)}=\sum_{\nu}C(1\nu
L+\lambda M-\nu|LM)Y_{L+\lambda M-\nu}(\hat{R})\m \xi_\nu\,\;,
\label{bm_5} \ee 
with $\m \xi_\nu$ being  three basic unit
vectors,
 $L$ --- the multipole order of the transition, and 
$m$ --- the corresponding magnetic quantum number. Below, we keep
the name of the transition nuclear current for its radial
component $J(R)$. This can be related to the  radiative amplitude
of the nuclear transition.
Expressing the latter as follows 
\be
F_\gamma=-i\int J^\mu(\m R)A_\mu^*(\m R)\,d^3R\,\;, 
\ee 
and
substituting the expression for the vector potential  in the case
of transitions of the magnetic type \cite{rose,ahi} \be \m
A^{(LM)}=-2\sqrt{\omega}j_L(\omega R)T_{LM}^{(0)} (\hat R)\,, \ee
one arrives at the equation relating the transition nuclear
current with the reduced transition magnetic momentum (or amplitude)
of the nuclear transition \cite{BM} 
\be \langle RJ(R)\rangle
\equiv \int_0^\infty J(R)R^3\ dR = \sqrt{(L+1)/L} \langle
I_2||{\cal M}(ML)||I_1\rangle \,. \label{mg} \ee 
Now, using in
(\ref{mg}) the definition of the nuclear magnetic moment $\mu$
\cite{BM} \be \frac{e\hbar}{2M_pc}\mu = \sqrt{\frac{4\pi}3}
\frac{C(II10|II)}{\sqrt{2I+1}} \langle I||{\cal M}(M1)||I\rangle
\,, \label{mm} 
\ee 
one arrives at the expression 
\be \langle R
J(R)\rangle = \sqrt{\frac3{2\pi}(2I+1)(I+1)/I}
\frac{e\hbar}{2M_pc}\mu    \,. \label{m1}
\ee
 Relations (\ref{mg})
-- (\ref{m1}) show that $RJ(R)$ can be treated as the radial distribution of
the magnetic currents over the nucleus. Therefore, baring in mind
application for studying nuclear structure, one can advance the
past expression to 
\be \langle R^{j+} J(R)\rangle =
\sqrt{\frac3{2\pi}(2I+1)(I+1)/I}
\frac{e\hbar}{2M_pc}\langle\mu\rangle_j  \,, 
\ee 
where $M_p$ is
the proton mass, $\langle\mu\rangle_j\equiv (R_j)^j$ --- the
$j$-th moment of the radial distribution of magnetization over the
nuclear volume.

    Energy shift of the state with the total angular momentum $F$ and
its projection $M$, nuclear and atomic momenta $I$ and $j$,
respectively, can be viewed as an elastic amplitude of the $M1$
internal conversion (IC) transition \cite{book,echa,HF}. The  amplitude is defined in terms of the
interaction of the nuclear and electronic transition currents: 
\be
F_c = \int  \left(\m J(\m R) \m s(\m r)\right) {\cal D}( |\m r -
\m R|) \ d^3r \ d^3R\,, \label{c1} \ee with the photon propagator
\be {\cal D}( |\m r - \m R|) = e^{|\m r - \m R|} / |\m r - \m R| =
4\pi i \omega \sum_{l=0}^\infty j_l(\omega r_<) h_l(\omega r_>)
Y^\star_{lm}(\hat R)  Y_{lm}(\hat r)\,,
    \label{fp} \ee 
and  the electron current
 \be \m s(\m r) = e\bar \psi \m \gamma \psi \equiv 4e\kappa \m j(\m
r)\,.    \ee 
In (\ref{fp}), $r_<$ ($r_>$) designates
the smaller (larger) of the $r$ or $R$, and $\hat r$ stands for
$\m r / r$. In turn, $j_L (\omega R)$, $h_L (\omega R)$ are the
spherical Bessel and Hankel functions, respectively. 
In the case of a point-like
nucleus,  $r_< = r$, $r_> = R$. These relations define the non-penetration (NP) model in the internal conversion theory. Within the NP model,  interaction
(\ref{c1}) with (\ref{fp}) is fully factorized with respect to the nuclear and electronic variables. The whole transition amplitude factorizes into the 
radiative nuclear amplitude and the remaining  factor, independent of the nuclear variables. The latter just defines 
the ICC \cite{rose,ahi}. With the account of the finite nuclear size,
such a  factorization may only be achieved within the framework of the nuclear models, among which the most widely used became the models of surface (SC) and volume (VC) transition nuclear currents. On the physical ground, the SC model is more justified by the Pauli principle, which to a greater extent prohibits motion of the internal nucleons. For this reason, the SC model is the cornerstone of known tables \cite{RAINE}. 
It recommended itself as working well in the internal conversion theory. 
Difference of experimental ICC from their tabular values is called as anomalies. The anomalies are thus related to the effects of penetration of electrons into the nuclear area. We see that the values of thus determined anomalies depend on the nuclear model. 

       The diagonal matrix element of the IC interaction Hamiltonian $H_c(L\mu)$  defines the HFS. $H_c(L\mu)$  is spherical tensor of rank $L$, $L$, $\mu$ being the multipole order and its magnetic quantum number.  
In order to maintain succession
with the internal conversion theory and previous papers
\cite{HF,book,echa}, we will consider formulae in  general case,
and in the end we will pass to the mathematical limit.

    Generally, amplitude (\ref{c1}) can be considered as a matrix element  of $H_c(L\mu)$:
\be F_c = \langle b|H_c(L\mu)|a\rangle\,, \label{mec} \ee 
where
$|a\rangle$, $|b\rangle$ are the are the wavefunctions of the
initial and final states of the atom. In the IC theory, they are
usually characterized with the quantum numbers $|IMjm\rangle$ of
the nuclear and electronic spins and their projections on the
quantization axis. On the other hand, the hyperfine shift $S_{FIj}$ is determined by the diagonal matrix element in
the $|FMIj\rangle$ representation, with $F$ and $M$ being the total
angular momentum of the atom and its projection. By making use of
the Wigner-Eckart theorem,  after 
straightforward
 algebra, the energy shift can be represented as follows \cite{wars}: 
\be S_{FIj} = \sum_\mu\langle
FMIj|H_c(L\mu)| FMIj\rangle = (-1)^{F-I-j} W\{IIjj;LF\} \langle
Ij||H_c(L)|| Ij\rangle  \,. \label{eqm}\ee The reduced two-bar
matrix element in eq. (\ref{eqm}) is defined as follows: \bea
\langle I_2M_2j_2m_2|H_c(LM)|I_1M_1j_1m_1\rangle =
\\
=
\frac{C(I_2M_2LM|I_1M_1)C(j_1m_1LM|j_2m_2)}{\sqrt{(2I_1+1)(2j_2+1)}}
\langle I_2j_2||H_c(L)||I_1j_1\rangle \eea

Recoupling the angular moments in (\ref{c1}), (\ref{fp}), one can
separate the angular variables \cite{rose,ahi} \be
\begin{split}
\frac{e^{i\omega|\m{r}-\m R|}}{|\m r-\m R|} \left(J(\m R) \m j(\m
r)\right)=4\pi i\omega\sum_{L\lambda M} \left(\m J\m
T_{LM}^{(\lambda)*}(\hat{\m R})\right) \\ \times \left(\m j\m
T_{LM}^{(\lambda)}(\m{\hat r})\right) j_{L+\lambda}(\omega r_<)
h_{L+\lambda}(\omega r_>)\,.    \label{bm_4}
\end{split}
\ee In the case of transitions of the magnetic type, the terms in
(\ref{bm_4}) with $\lambda$ = 0 are relevant. Using definition
(\ref{2}) and  integrating over the electronic variables, as shown
in \cite {rose, ahi}, one arrives at the following expression:
\bea \langle Ij||H_c(L)|| Ij\rangle =
\frac{16\alpha\pi\omega\kappa}{\sqrt{L(L+1)}}i^{l-l^\prime
+1}(-1)^{l^\prime-j-L-\oh} W\{l^\prime jlj;\oh L\} G\,,  \\
G = \int_0^\infty J(R) j(r) j_L(\omega r_<) h_L(\omega r_>) \ R^2
dR \ r^2  dr\,. \label{ME} \eea
 Inserting (\ref{ME}) into 
(\ref{eqm}), one arrives at the following expression for the
hyperfine splitting $w$ of the states with $F=I+j$ and $F=I-j$:
\be w=-4e\omega\kappa\sqrt{\frac{6\pi j(2I+1)}{I(I+1)(j+1)}} G \,.
\label{whf} \ee

    The fact that the hyperfine shift is generically related to the conversion 
amplitude is manifested in the limiting relation between the
formulae for the ICC at $\omega\rightarrow 0$ and for the HFS 
 \cite{book,echa}
\begin{multline}
S_{FIj}=(-1)^{j+I-F}W(IIjj;LF)\frac{2}{(2L+1)!!}
\left(1+\frac{m}{M}\right)^{-3} \times 
 \\ {\times \lim_{\omega\rightarrow 0}
\sqrt{\omega^{2L+L}\alpha_\partial (M1)(2j+1)(2L+1)(L+1)/L}
}\times
\\ \times \langle I\|{\cal M}(M1)\|I\rangle\,\;, \label{hf8}
\end{multline}
with $L=1$, where $m$ and $M$ are the electron and nuclear masses,
respectively. In (\ref{hf8}), $\alpha_\partial (\tau l)$ is the
analogue of traditional ICC $\alpha (\tau l)$ extended to the case where the conversion electron occupies a discrete electronic state \cite{mux,atalah}. It has
dimension of energy. Note  that  eq.(\ref{hf8}) is exact in the
sense that its dependence on the nuclear model can be fully
related to the way of calculation of ICC. In the SC model, the magnetic current reads as follows \cite{RAINE, SC}: \be
J_{SC}(R)=D\delta(R-R_0)    \,, \label{SC}  \ee $R_0$ being radius
of the magnetic current. With $R_0$ = 0, this model includes the
NP model. The constant $D$ as determined from the normalization
condition (\ref{m1}) reads as follows: \be D =
\sqrt{\frac3{2\pi}(2I+1)(I+1)/I} \frac{e\hbar}{2M_pc}\mu /R_0^3
\,. \label{mD}\ee Then eq. (\ref{ME}) is factorized, resulting in
\bea w =\frac{2(2I+1)}{I(j+1)} e\omega^2 \kappa
\frac{e\hbar}{2M_pc} \mu i{\cal F} \,, \label{wsc}\eea where
 $\kappa$  is the
relativistic quantum number, and the radial electronic matrix
element \be {\cal{F}} = \int_0^\infty j(r) X_L(\omega R) r^2 \
dr\,. \label{elr} \ee Eq. (\ref{wsc}) can also be obtained from
(\ref{hf8}). In (\ref{elr}), the electronic transition current \be
j(r) = g(r)f(r) \,. \label{elc} \ee  
	In (\ref{elc}), $g(r)$, $f(r)$ are the 
 large and small radial Dirac wavefunctions of the electron.
Furthermore, $X_L(\omega R)$ is the potential of the electronic
transition. $X_L = h_L(\omega R)$ within the NP model. In the SC
model it reads \be X_L(\omega R) = [h_L(\omega R_0) / j_L (\omega
R_0)] j_L (\omega R)\,. \label{SC} \ee The related ICC in eq.
(\ref{hf8}) reads as follows: \bea \alpha_d
(\tau L) = |Q^{(L)} {\cal F}|^2\,, \\
 Q^{(L)} = -4\kappa\sqrt{\frac{\alpha \pi
\omega}{L(L+1)}} C(j-\oh L0|j-\oh)\,. \eea

\section{The formalism}

In  any realistic nuclear model, after separation of the angular
variables, the electronic and nuclear variables are mixed in the
radial conversion matrix element (\ref{ME}).   For the purpose of
separation of the penetration effects, let us express the nuclear transition current
in general form as \be J( R) = J_{SC}( R) + [J( R) - J_{SC}( R)]
\equiv J_{SC}( R) + J_p ( R) \label{6} \ee and, correspondingly,
the matrix element (\ref{ME}) --- in the form \be G = G_{SC} + G_p\,. \ee Here
$G_{SC}$ is the factorized SC model contribution, and $G_p$ bares
information on the penetration effects.  It reads as \bea G_p =
\int_0^\infty J_p ( R)  R^2 dR \ [ j_L(\omega R)
    \int_R^\infty j(r) h_L(\omega r) r^2dr  + \nonumber\\+
     h_L(\omega R)  \int_0^R  j(r) j_L(\omega r) r^2dr] =
\int_0^\infty J_p ( R) Q(R) \ R^2 dR \,,   \label{7} \\ Q(R) =
\int_0^R \left[ h_L(\omega R)  j_L(\omega r)  - j_L(\omega R)
h_L(\omega r)\right]\ j(r) r^2dr  \,.    \label{7a} \eea

    In the nuclear area, the electronic wavefunctions are well
represented by the Taylor series \cite{tay,RAINE}: \bea g(r) =
a_0+a_2r^2+\ldots, \qquad f(r) = b_1r+b_3r^3+\ldots   \qquad
\text{for the $s$ states}\,,
\label{tay1}    \\
g(r) = a_1r+a_3r^3+\ldots, \qquad f(r) = b_0+b_2r^2+\ldots \qquad
\text{for the $p_\oh$ states}\,.    \label{tay2}     \eea

By making use of (\ref{tay1}), (\ref{tay2}) and asymptotic
expansions for the spherical Bessel and Hankel functions, the
electronic current in (\ref{7}) can be put down as follows: \be
j(r) = c_1r + c_3 r^3 + \ldots \,. \label{jas} \ee
As a result, we arrive at the series expansion for $Q(R)$ and,
finally,
 \be G=\frac1\omega
\sqrt{\frac3{2\pi}(2I+1)(I+1)/I} \frac{e\hbar}{2M_pc}\mu
\Bigl[\frac{\omega^2}3 {\cal F} +
\sum_{i=2,4,\ldots}\frac{c_i}{i(i+3)}\left(R_0^{i}-
(R_{i})^{i}\right)\Bigr]\,.\label{eqG}
\ee
   Therefore, by comparing experimental HFS with the theory for an
atomic level, we obtain the root mean square radius $R_2$ of the nuclear magnetization, and can
unambiguously predict the HFS value for all the other $s$- and
$p_\oh$ levels. This parameter can associate with the effective nuclear radius which is usually extracted from the analysis of the data concerning the nuclear hyperfine anomalies.  Furthermore,  comparing experiment with theory for
two levels, one can also find the fourth  moment of the nuclear
magnetization distribution and, correspondingly, predict more
precisely the HFS value for the other levels, etc. Analysis of $n$
levels provides us with the even nuclear magnetization moments up
to the $2n$-th one, inclusively.

\section{Results}

    Atomic calculations were performed by means of the package of
computer codes RAINE \cite{RAINE}. 
 Fermi nuclear charge
distribution was supposed, with typical parameters.  Higher order
QED corrections were taken into account. A typical value of $R_0$
= 7.121 fm was  adopted for the radius of the nuclear magnetic
current. The vacuum polarization potential and
the electron selfenergy correction were allowed for  as suggested in refs. \cite{VP14,VP15}. As a result, for the $1s$ state, the value was obtained
$w_h$ = 5.107 eV, in coincidence with the previous calculation
\cite{HF}. Theoretical calculation should then be compared to the
experimental value. For a long time, the value of 5.0841$_8$~eV
was accepted \cite{klaft}. A somewhat lower value of
5.0863$_{11}$~eV \cite{exli}, though outside  the range of the
previous error bars, was obtained recently. Furthermore, the
 wavefunctions, calculated in this way, were used in order to fit  the experimental data.
To this end, we incorporated into the calculation the latest QED
correction \cite{se2,pers2} $\Delta w_{QED}$ = $-$0.0268 eV.
Being added to the basic ``Coulomb" value of $w^C$ = 5.0825 eV of the
bare nucleus, calculated by us taking into account pure electrostatic electron interaction with the nucleus, this results in the calculated HFS
value of $w_{th}$ = 5.0558 eV, which is by 0.6\% lower that the
experimental value \cite{exli}. Then, dealing with the difference between
the theory and the latter experimental value as shown previously, we
achieve consensus with the experimental value for this level
 with the root-mean-square radius  of
the nuclear magnetization $R_2^{1s}$ = 6.207 fm. This is the only
parameter which we can determine by comparison of theory with
experiment for one level. We dwell on these details in order to
show a possible range of scatter of the parameters and sensitivity
of the model to them.

 At the time being, there are  available  data concerning
 the HFS values for the upper
states.  The latest one obtained in the Experimental Storage Ring
at GSI is 0.79750(18) eV \cite{exli}. This is noticeably lower in 
comparison with  the earlier measurement at the Lawrence Livermore
National Laboratory \cite{exca} of 0.820$\pm$0.026 eV. Regarding
theory, our calculation for a bare
nucleus, with the Fermi charge distribution with usual parameters,
and $R_0$ = 7.121 fm yielded in the $w_C^{2s} = 0.826$~eV. QED
correction may be introduced according to \cite{shab2} as $\Delta
E_{QED}^{2s} = -0.005$~eV.
The most essential difference with the analysis of the
$1s$ case arises in the necessity of account of the
electron-electron interaction $\Delta E_{e-e}^{2s}$. In the zeroth
approximation, we allowed for that through the self-consistent
Dirac-Fock method \cite{RAINE}. As a result, we arrived at the value of $\Delta E_{e-e}^{2s}$ = $-$0.038 eV. This a little exceeds the  calculation \cite{shab2}, where all the terms  up to the second order in the $\alpha Z$ perturbation series were taken into account, and the electron
screening correction of $\Delta E_{e-e}^{2s}$ = $-$0.030 eV was 
obtained 
If we  rely on
the latter value, we arrive at the HFS of $w_{th}^{2s}$ = 0.791 eV.
Fit to the experimental value of the HFS for this one level only,  results then in
 the root-mean-square
radius of the nuclear currents of $R_2^{2s}$ = 6.275 fm. We see
that $R_2^{1s} \neq R_2^{2s}$, as we neglect the remaining terms
in eq. (\ref{eqG}) including $R_4$ and higher moments. We then
found that with account of the two radii, $R_2$ and $R_4$, a
better fitting the multipole moments of the magnetization  is
achieved if a value of $w_{th}^{2s}$ = 0.792 eV is used. Then
solving the system of coupled algebraic equations (\ref{eqG}) for
the $1s$- (in the H-like ions) and $2s$- (in the Li-like ions)
states results in the reasonably close value of $R_2$ = 6.118 fm,
and $R_4$ = 6.760 fm. Note that the radii obtained satisfy a
plausible relation $R_2 < R_4$.

	As a test for the self-consistency of the method, we repeated the
same calculations with another value of $R_0$ = 6.2 fm, which is
closer to the $R_2$ obtained and therefore, should be more
realistic, and provide with a better convergence of the series
(\ref{eqG}), as explained above. Indeed, we obtained as the
starting values of $w_{th}^{1s}$ = 5.079 eV, $w_{th}^{2s}$ = 0.796
eV, which values are considerably closer to the experimental ones.
As expected, by solving the system of two coupled equations
(\ref{eqG}) we arrived at the same value of $R_2$ = 6.12 fm, and
slightly corrected value of $R_4$ = 6.78 fm, differing by 0.3\%.
Now this was achieved already without additional varying the
$w_{th}^{2s}$ value. The scheme of calculation is illustrated in Table 1.
\begin{table}
\caption{\footnotesize Representative results of calculation of
the $R_2$ and $R_4$ magnetic radii. In the first column, the
levels used for solving eqs. (\ref{eqG}) are indicated. With both the levels involved in the analysis, the resulting $R_2$ and $R_4$ parameters hold in spite of essentially different starting values of the SC radius $R_0$. }
\begin{center}
\begin{tabular}{c|c|c|c}
level&$R_0$, fm & $R_2$, fm & $R_4$, fm \\
\hline
$1s$ & 7.12 & 6.21 & --- \\
$2s$ & 7.12 & 6.27 & --- \\
both & 7.12 & 6.12 & 6.76 \\
both & 6.20 & 6.12 & 6.78 \\
\hline
\end{tabular}
\end{center}
\label{tab}
\end{table}

 Let us see how this works for the $2p_{\oh}$ state. Direct data concerning the HFS value for the upper
states, are not available at the time being, to the best of our
knowledge. 
Proceeding in the way similar to the $2s$ case above, we obtained the Coulomb base value of $w^C$ = 0.28509 eV for the $2p_{\oh}$ state. Account of the $R_2$- and $R_4$-terms in eq. (\ref{eqG}) lowers this value to 0.28498 eV. Allowing for the correction $\Delta w_{e-e}$ = $-$27.19 meV, owing to the electron-electron interaction, and the QED correction $\Delta w_{QED}$ = $-$0.26 meV \cite{sha2p}, we arrive at the resulting value of $w_{2p}$ = 0.25753 eV. This may be compared to the theoretical value of 257.84(5) meV predicted in Ref. \cite{sha2p}.

\section{The Bohr-Weisskopf effect}

	In a series of papers, it was proposed to cancel the BW effect in
linear combination of the HFS's of the two levels (specific difference) with
a parameter $\zeta$ (\cite{shab1,shab2} and refs. cited therein):  
\be
\Delta^\prime w = w^{2s} - \zeta w^{1s}\,,	\label{sdf} \ee 
with
the $\zeta$ value of 0.16886. 	 However, a model-dependent character of
such relations was shown above. Eq. (\ref{sdf}) is derived from
the conventional Weisskopf model with constant nucleon
wavefunctions inside the nuclei. Generally, the specific differences (\ref{sdf}) are
calculated with the same accuracy like each of the terms, that is
$\sim$1\%. It is worthy of noting here that the exact BW value is
unobservable experimentally: this is a purely estimated value. The
ultimate definition of the BW effect can be formulated as the difference
between the experimental value of HFS and that calculated within the
NP model: \be
 b_\kappa = (w_{NP} - w_{exp}) / w_{NP} \,. \label{bw}
\ee The latter model would unambiguously and adequately predict
the anticipated effect for the  corresponding point-like nucleus.
This model excludes influence of nuclear dynamics. At the same time, in this
definition static nuclear properties like charge distribution
influence the calculated NP value, together with QED corrections.
We will not further separate out these effects in the definition
(\ref{bw}). We summarize that the calculated NP values are 5.162,
0.811, and 0.25828 eV for the HFS values of the $1s$-, $2s$- and
$2p_\oh$ levels, respectively. These result in the BW effect of 1.73, 1.47 and 0.29 percent, respectively. It is these values of the BW effect which predetermines the 
difference between various models. We recall that the static nuclear properties contribute at the level of up to
10\% to the ICC values  in the case of heavy nuclei (e.g.
\cite{HI} and refs. therein), that is even more than  the BW
effect (\ref{bw}).

	 Our above fit to the data on the HFS of the both $1s$- and $2s$
levels results in the $\zeta$ value of 0.1746, that is 3\% higher than that cited above.
This discrepancy is just what was fairly expected in the light of what was  said above. We conclude that 
parameter $\zeta$ in  the specific difference (\ref{sdf}) is neither observable, nor model-independent enough, to be predicted up to five digits by making use of only one or two experimental HFS values for the element.

\section{Discussion}

	Consrcutive theory of HFS, including the description of the anomalies caused by the Bohr---Weisskopf effect, is presented previously. That is founded on the classical theory of anomalous internal conversion.  	
First, the theory shows the kind of information which can be obtained from studies of HFS'. This is even multipole moments of the
nuclear magnetization distribution. They can be extracted by means
of   solving  the system of coupled equations (\ref{eqG}) for 
 several electronic configurations, for which the experimental HFS$^\prime$ are available.
 Analysis of the
data on the HFS in the H-like and Li-like ions of $^{209}$Bi resulted in 
the second- and, for the first time in internal conversion theory, the forth-momentum  radii: $R_2$ = 6.12
fm, and $R_4$ = 6.78 fm. 
Dependence of the HFS on the $R_4$ and higher moments of the magnetization distribution turns out to be a specific feature  of HFS. Traditional IC theory assumes  that there is only one parameter, $R_2$,  which can be determined from comparison of theory with experiment. The same situation ican be noted in the contemporary analysis of the anomalies in the HFS, observed within the isotopic chains: usually the only  nuclear parameter, equivalent to $R_2$, is discussed (e.g., \cite{barza,kara}).
Therefore, dependence of HFS on the higher magnetization distribution moments, demonstrated above,  is an extension of the internal conversion theory. Furthermore, 
determination of higher radii from experiment is impossible in the absence of precision atomic calculations with proper account of higher order QED corrections. 
Hence, the above way of analysis simultaneously offers a critical test of QED, otherwise impossible. The stringency of the test is only limited by the number of the electronic configurations of the atom of the same element for which data on the HFS are available. Our fit confirmed validity of the QED corrections calculated previously, within the framework of the latest experimental data.

As we saw above, the corrections at the level of $\sim$1 meV
to the HFS of the $2s$ level were crucial for treating the BW
effect, which comprises $\sim$0.1\% of the shift. This clarifies the 
sensitivity of the method to the parameters. The $R_4$ value was
obtained within the accuracy of about 0.3\%, and the $R_2$ value was
most stable against fitting, not changed at all. This estimation shows 
more definitely the limits of the theory, as well as its abilities and prospects concerning future experimental data while they will be available.

	 These conclusions are derived by means
of consecutive treatment of the BW effect, instead of fighting it, e.g.
through combining the specific differences. The principal defect of the latter method is the loss of information concerning the nuclear structure, as this information is just conveyed {\it via} the BW effect. Another defect is that the recipe itself is erroneous, being founded on a model consideration. This means that for more than well two decades the research in this field  has been actually going in the wrong direction, when fighting the BW effect instead of using it fruitfully for retrieving information from the data. Narrowing of the theme is also redicing the discussion about the nuclear anomalies to the single parameter --- the effective nuclear radius, losing sight of $R_4$ and other moments. 

	Prediction is also made for the HFS of the $L2$-electron, in which  case data are not available up to date. Future experiments should manifest real relation between the discussed factors.

	To summarize, the new nuclear-model independent way of treating the HFS is proposed, proceeding from the theory of anomalies which were first observed in ICC. The way is based on simultaneous analysis 
of experimental data concerning HFS for several electronic configurations
 of the atom. The method can be used as a stringent test of QED. Consequently, further development of the
experimental basic research is needed, aimed at measuring the HFS on
ions with various few-electronic configurations with a small
number of electrons, where calculation of the electronic interaction is
more feasible. For this purpose, the storage rings are suitable
which are available or being in reconstruction e.g. at the Lawrence Livermore National Laboratory, GSI
Darmstadt, Lanzhou in China.  From the theoretical point of view,
the QED corrections, nuclear recoil effect are of great importance
for treating future data.  A more consecutive way of incorporation of
these effects into the calculation of the atomic wavefunctions
looks quite feasible and should be pursued in further
invest1igation.	\\

\qquad
	The authors would like to acknowledge many detailed discussions of the topic with D. Glazov. They are thankful to A.E.Barzakh, L.N.Labzovsky, Yu.Litvinov, V.M.Shabaev and L.F.Vitushkin for  fruitful discussions and helpful comments.

\end{document}